\documentstyle[aps,twocolumn,epsfig,floats,eqsecnum]{revtex}

\newcommand{\eins}{{\openone}}
\newcommand{\spur}{\text{tr}\,}
\newcommand{\imagteil}{\text{Im}\,}
\newcommand{\veps}{\varepsilon}
\newcommand{\pvec}{{\bf \Psi}}
\newcommand{\lmat}{\hat L}

\newcommand{\sigmax}{\hat \sigma_x}

\newcommand{\sigmaz}{\hat \sigma_z}
\newcommand{\umat}{\hat u}
\newcommand{\samat}{\hat S}

\newcommand{\diag}{\text{diag}}
\newcommand{\rmat}{\hat R}
\newcommand{\str}{\text{Str}}
\newcommand{\sdet}{\text{Sdet}}
\newcommand{\tmat}{\hat T}
\newcommand{\qmat}{\hat Q}
\newcommand{\hmat}{{\cal H}}
\newcommand{\erf}{\text{erf}\,}
\begin{document}

\draft

\title{Quantum limit of the laser linewidth in chaotic cavities and
statistics of residues of scattering matrix poles}
\author{H. Schomerus$^{\rm a}$,
K. M. Frahm$^{\rm b}$,  M. Patra$^{\rm a}$,
and C. W. J. Beenakker$^{\rm a}$}
\address{{}$^{\rm a}$Instituut-Lorentz, Universiteit Leiden,
P.\,O.~Box 9506, NL-2300 RA Leiden,
	The Netherlands
	\\
	{}$^{\rm b}$Laboratoire de Physique Quantique, UMR 5626 du CNRS,
			 Universit\'e{} Paul Sabatier, F-31062 Toulouse Cedex 4, France
	}
\date{\today}

\twocolumn[
\widetext
\begin{@twocolumnfalse}

\maketitle

\begin{abstract}
The quantum-limited linewidth of a laser cavity is
enhanced above the Schawlow-Townes value by the
Petermann factor $K$,
due to the non-orthogonality of the cavity modes.
We derive the relation between the Petermann factor and the
residues of poles of the scattering matrix and
investigate the statistical properties of the Petermann factor
for cavities in which the radiation is scattered chaotically.
For a single scattering channel
we determine the complete probability distribution
of $K$ and find that
the average Petermann factor
$\langle K \rangle$
depends non-analytically
on the area of the opening, and
greatly exceeds the most probable value.
For an arbitrary number $N$ of scattering channels we calculate
$\langle K\rangle$
as a function of the
decay rate $\Gamma$
of the lasing mode.
We find for $N\gg 1$ that for typical values of $\Gamma$
the average Petermann factor
$\langle K\rangle\propto \sqrt{N}\gg 1$
is parametrically larger than unity.

\end{abstract}

\pacs{PACS: 42.50.Lc, 42.50.Ar, 42.60.Da}

\vspace{0.5cm}

\narrowtext
\end{@twocolumnfalse}
]
\narrowtext

\section{Introduction}
Laser action selects a mode in  a cavity and enhances the output
intensity in this mode by a non-linear feedback mechanism.
Vacuum fluctuations of the
electromagnetic field ultimately limit the narrowing of the emission
spectrum~\cite{schawlow:58a}. The quantum-limited linewidth, or
Schawlow-Townes linewidth,
\begin{equation}
	\delta\omega_{\rm ST}^{} = \case{1}{2} \Gamma^2 / I,
\label{startgl}
\end{equation}
is proportional to the square of the decay rate $\Gamma$ of the lasing
cavity mode
and inversely proportional to the output power $I$ (in
units of photons/s). This is a lower bound for the
linewidth when 
$\Gamma$ is much less than the linewidth
of the atomic transition and when the lower level of the transition is
unoccupied.
Many years after the work of Schawlow and Townes it was
realized~\cite{petermann:79a,siegman:89} that the true
fundamental limit is larger than Eq.\ (\ref{startgl}) by a factor $K$
that characterizes the non-orthogonality of the cavity modes. This
excess noise factor, or Petermann factor, has generated an extensive
literature
\cite{cheng:96a,eijkelenborg:96a,brunel:97a,grangier:98a,siegman:98a}.

Apart from its importance for cavity lasers, the Petermann factor is of
fundamental significance in the more general context of scattering theory.
A lasing cavity mode is associated with a pole of the scattering matrix
in the complex frequency plane. 
We will show that the Petermann factor
is proportional to the squared modulus of the residue of
this pole.
Poles of the scattering matrix
also determine the position and height of resonances of nuclei, atoms, and
molecules \cite{MW}.
Powerful numerical tools that give access to
poles even deep in
the complex plane have been developed recently \cite{harminv}.
They can be used to determine the residues of the poles as well.
Our work is of relevance for these
more general studies, beyond the original application to cavity lasers.

\begin{figure}
\center{\epsfig{file=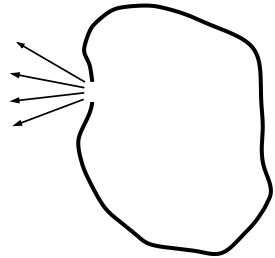,width=1.4in}}
\medskip
\caption{Chaotic cavity that radiates light from a small opening.
}
\label{fig:cavity}
\end{figure}

Existing theories of the Petermann factor deal with cavities in which
the scattering is essentially one-dimensional, because the geometry has a
high degree of symmetry. For such cavities
the framework of ray optics provides a simple way to solve the problem in
a good approximation \cite{eijkelenborg:96a}. 
This approach breaks down if the light propagation in the cavity
becomes chaotic, either because of 
an irregular shape of the boundaries
(like for the cavity depicted in Fig.\ \ref{fig:cavity})
or because of randomly placed scatterers.
The method of random-matrix theory is well-suited for such
chaotic cavities~\cite{haake:91,mehta:90}.
Instead of considering a single cavity, one studies an ensemble of
cavities with small variations in shape and size, or position of the
scatterers. The distribution of the scattering matrix in this ensemble
is known. Recent work has provided a detailed knowledge on the
statistics of the poles
\cite{sokolov:88,fyodorov:96a,fyodorov:97a,fyodorov:99a}.
Much less is known about the residues \cite{janik:99a,chalker:98a}.
In this work we fill the remaining gap to a considerable extent.

The outline of this paper is as follows.
In Section \ref{sec:qoptics}
we derive the 
connection between the Petermann factor and the residue of the pole of the
lasing mode. The residue in turn is seen to be characteristic 
for the degree of nonorthogonality of the modes. In this way we make
contact with the existing literature on the Petermann factor
\cite{grangier:98a,siegman:98a}.

In Section \ref{sec:onechannel} we study the single-channel case
of a scalar scattering matrix. 
This applies to a cavity that is coupled to the outside via 
a small opening of area ${\cal A}\lesssim 
\lambda^2/2\pi$ (with $\lambda$ the wavelength of the lasing mode).
For preserved time-reversal symmetry (the relevant case in optics)
we find that the ensemble average of $K-1$ depends {\em
non-analytically} $\propto T \ln T^{-1}$ on the transmission probability $T$
through the opening, so that it is beyond the reach of perturbation
theory even if $T\ll 1$. We present a complete resummation of the
perturbation series that overcomes this obstacle.
We derive the conditional distribution $P(K)$ of the Petermann factor 
at a given decay rate $\Gamma$ of the lasing mode, valid for any 
value of $T$.
The most probable value of $K-1$ is $\propto T$, hence it is
parametrically smaller than the average.

In a cavity with such a small opening the deviations of $K$ from
unity are very small. For larger deviations we study, in
Section \ref{sec:manychannels}, the
multi-channel case of an $N\times N$ scattering matrix, which corresponds
to an opening of area ${\cal A}\approx N \lambda^2/2\pi$.
The lasing mode acquires a decay rate $\Gamma$ of order
$\Gamma_0=N T\Delta/2\pi $ (with $\Delta$ the mean spacing of the cavity
modes).
We compute the mean Petermann factor as a function
of $\Gamma$ for broken time-reversal symmetry, which is technically
simpler than the case of preserved time-reversal symmetry, but
qualitatively similar.
We find a parametrically
large mean Petermann factor $K\propto \sqrt{N}$.

Our conclusions are given in Section \ref{sec:conclusions}.
The main results of Sections \ref{sec:onechannel} and
\ref{sec:manychannels} have been reported 
in Refs.\ \cite{patra:99} and \cite{frahm:99}, respectively.

\section{Relationship between Petermann factor and residue}
\label{sec:qoptics}

Modes of a closed cavity, in the absence of
absorption or amplification, are eigenvalues $\omega_n$
of a Hermitian operator
$H$. This operator can be chosen
real if the system possesses time-reversal symmetry
(symmetry index $\beta=1$), otherwise it is
complex ($\beta=2$).
For a chaotic cavity, $H$ can be modeled by an $M\times M$
Hermitian matrix with independent Gaussian distributed elements,
\begin{equation}
P(H)\propto
\exp\left[-\frac{\beta M}{4\mu^2}\spur H^2\right].
\label{eq:hdist}
\end{equation}
(For $\beta=1$ (2), this is the Gaussian orthogonal (unitary)
ensemble~\cite{mehta:90}.)
The mean density of eigenvalues is the Wigner semicircle 
\begin{equation}
\rho(\omega)=\frac{M}{2\pi\mu^2}\sqrt{4\mu^2-\omega^2}.
\label{eq:semicircle}
\end{equation}
The mean mode spacing at the center $\omega=0$ is $\Delta=\pi\mu/M$.
(The
limit $M\to\infty$ at fixed spacing $\Delta$ of the modes is taken at
the end of the calculation.)

A small opening in the cavity is described by
a real, non-random $M\times N$ coupling matrix $W$, with $N$ the
number of scattering channels transmitted through the opening. (For an
opening of area ${\cal A}$, $N\simeq 2\pi {\cal A}/\lambda^2$
at wavelength
$\lambda$.)
Modes of the open
cavity are complex eigenvalues
(with negative imaginary part) of the non-Hermitian
matrix 
\begin{equation}
\hmat =H - i\pi W W^\dagger
.
\label{eq:hmat}
\end{equation}

In absence of amplification or absorption,
the scattering matrix $S$ at frequency
$\omega$ is related to $\hmat$ by~\cite{MW,verbaarschot:85a}
\begin{equation}
	S = \openone - 2 \pi i W^\dagger ( \omega - \hmat )^{-1} W .
	\label{Smatrix0}
\end{equation}
The scattering matrix is a unitary (and symmetric, for $\beta=1$)
random $N\times N$ matrix, with poles at the
eigenvalues of $\hmat$. It
enters the input-output relation
\begin{equation}
a_m^{\rm out}(\omega)=\sum_{n=1}^N S_{mn}(\omega)a_n^{\rm in}(\omega)
,
\end{equation}
which relates the annihilation operators $a_m^{\rm out}$ 
of the scattering states that leave the cavity to the
annihilation operators $a_n^{\rm in}$ of states that
enter the cavity. The indices $n$, $m$ label the scattering channels.

We now assume that the cavity is filled with a homogeneous amplifying
medium (constant amplification rate $1/\tau_a$ over a large frequency
window $\Omega_a=L\Delta$, $L\gg N$).
This adds a term $i/2\tau_a$
to the eigenvalues, shifting them upwards towards the real axis.
The scattering matrix 
\begin{equation}
	S = \openone - 2 \pi i W^\dagger ( \omega - \hmat -i/2\tau_a )^{-1} W 
	\label{Smatrix}
\end{equation}
is then no longer unitary,
 and the input-output relation changes to
 \cite{jeffers,beenakker1998a} 
\begin{equation}
a_m^{\rm out}(\omega)=
\sum_{n=1}^N S_{mn}(\omega)a_n^{\rm in}(\omega)
+
\sum_{n=1}^N Q_{mn}(\omega)b_n^\dagger(\omega)
.
\end{equation}
All operators fulfill the canonical bosonic commutation relations
$[a_n(\omega),a_m^\dagger(\omega')]=\delta_{nm}\delta(\omega-\omega')$.
As a consequence,
\begin{equation}
Q(\omega)Q^\dagger(\omega) =
S(\omega)S^\dagger(\omega) -\eins.
\end{equation}
The operators $b$ describe the spontaneous
emission of photons in the cavity and have expectation value
\begin{equation}
\langle b_n^\dagger(\omega) b_m(\omega') \rangle=
\delta_{nm}\delta(\omega-\omega')f(\omega,T),
\end{equation}
with $f(\omega,T)=
[\exp(\hbar \omega/k_BT)-1]^{-1}$ the Bose-Einstein distribution function
at frequency $\omega$ and temperature $T$.

In the absence of external
illumination ($\langle a^{\dagger \rm in} a^{\rm in}\rangle=0$),
the photon current per frequency interval,
\begin{equation}
I(\omega)=\frac{1}{2\pi}
\sum_{m=1}^N\langle a_m^{{\rm out}\dagger}(\omega)a_m^{\rm out}(\omega)
\rangle
,
\end{equation}
is related 
to the scattering matrix by Kirchhoff's law
\cite{jeffers,beenakker1998a}
\begin{equation}
I(\omega)=f(\omega,T)\frac{1}{2\pi}\spur[\eins-S^\dagger(\omega)
S(\omega)].
\end{equation}
For $\omega$ near the laser
transition we may replace $f$ by the population inversion factor $N_{\rm
up}/(N_{\rm low}-N_{\rm up})$, where $N_{\rm up}$ and
$N_{\rm low}$ are the mean occupation numbers of the upper and lower
levels of the transition.
In this way the photon current
can be written in the form
\begin{equation}
I(\omega)=
\frac{1}{2\pi}\frac{N_{\rm
up}}{N_{\rm up}-N_{\rm low}}
\spur[S^\dagger(\omega) S(\omega)-\eins]
,
\end{equation}
that is suitable for an amplifying medium. (Alternatively, one can
associate a negative temperature to an amplifying medium.)

The
lasing mode is the eigenvalue $\Omega-i\Gamma/2$
closest to the real
axis, and the laser threshold is reached when the decay rate $\Gamma$
of this mode equals the amplification rate $1/\tau_a$.
Near the laser
threshold we need to retain only the contribution from the lasing mode
(say mode number $l$) to the scattering matrix~(\ref{Smatrix}),
\begin{equation}
	S_{nm} = - 2 \pi i \frac{(W^\dagger U)_{nl} (U^{-1} W)_{lm} }{\omega-\Omega
	+ i \Gamma/2-i/2\tau_a } ,
	\label{Smatrix2}
\end{equation}
where $U$ is the matrix of right eigenvectors of ${\cal H}$
(no summation over $l$ is
implied).
The photon current near threshold takes the form
\begin{equation}
I(\omega)=\frac{2\pi N_{\rm
up}}{N_{\rm up}-N_{\rm low}}
\frac{
 (U^\dagger W W^\dagger U)_{ll}
 (U^{-1 }WW^\dagger U^{-1 \dagger})_{ll}
}{ (\omega -\Omega)^2
	+  \frac{1}{4} (\Gamma -1/\tau_a)^2 }
	.
\end{equation}
This is a Lorentzian with full width at half maximum $\delta \omega=
\Gamma-1/\tau_a$.

The coupling matrix $W$ can be eliminated by writing
\begin{mathletters}
\begin{eqnarray}
-\pi (U^\dagger W W^\dagger U )_{ll}
&=&
\imagteil \left(U^\dagger {\cal H} U \right)_{ll}
\nonumber 
\\
&=&
-\frac{\Gamma}{2} ( U^\dagger U)_{ll}
,
\\
-\pi (U^{-1 }W W^\dagger U^{-1\dagger} )_{ll}
&=&
\imagteil \left(U^{-1 } {\cal H} U^{-1\dagger}\right)_{ll}
\nonumber
\\ &=&
-\frac{\Gamma}{2} (U^{-1 } U^{-1\dagger})_{ll}
.
\end{eqnarray}
\end{mathletters}%
The total output current is found by integrating over frequency,
\begin{equation}
I=\
(U^\dagger  U)_{ll} (U^{-1 } U^{-1\dagger})_{ll}
\frac{N_{\rm up}}{N_{\rm up}-N_{\rm low}}
\frac{\Gamma^2}{\delta\omega}.
\end{equation}
Comparison with the Schawlow-Townes value (\ref{startgl}) shows that
\begin{equation}
\delta\omega=2K\frac{N_{\rm up}}{N_{\rm up}-N_{\rm low}}
\delta\omega_{\rm ST}^{}
,
\end{equation}
where the Petermann factor $K$ is identified as 
\begin{equation}
K=
(U^\dagger  U)_{ll} (U^{-1 } U^{-1\dagger})_{ll}
\ge 1
.
\label{eq:Kgl1}
\end{equation}
For time-reversal symmetry, we can choose $U^{-1}=U^T$, 
and find $K=[(U U^\dagger )_{ll}]^2$.
The factor of $2$ in the relation between $\delta\omega$ and 
$\delta\omega_{\rm ST}^{}$ occurs because we have computed the laser linewidth
in the linear regime
just below the threshold, instead of far above the threshold.
The effect of the non-linearities above threshold is to suppress the amplitude
fluctuations
while leaving the phase fluctuations intact \cite{goldberg}, hence
the simple factor of two reduction of the linewidth.
The factor $N_{\rm up}/(N_{\rm up}-N_{\rm low})$ accounts for
the extra noise due to an incomplete population inversion.
The remaining factor $K$ is due to
the non-orthogonality of the cavity modes
\cite{siegman:89}, since $K=1$ if $U$ is unitary.

\section{Single scattering channel}
\label{sec:onechannel}

The relation (\ref{eq:Kgl1})
serves as the starting point for a calculation of the
statistics of the Petermann factor in an ensemble of chaotic cavities. 
In this section we
consider the case $N=1$ of a single scattering channel, for which the
coupling matrix
$W$ reduces to a vector $\vec{\alpha}=(W_{11}, W_{21},\ldots,W_{M1})$.
The magnitude $|\vec{\alpha}|^2=(M\Delta/\pi^2) w$,
where $w\in [0,1]$ is related to the
transmission probability $T$ of the single scattering
channel by $T=4 w(1+w)^{-2}$ \cite{BeenRMP}.
We assume a basis in which $H$ is diagonal 
(eigenvalues $\omega_q$,
right eigenvectors $|q\rangle$, left eigenvectors $\langle q|$).
In this basis the
entries $\alpha_q$ remain real for $\beta=1$, but become complex numbers
for $\beta=2$. Since the eigenvectors $|q\rangle$ point into random directions,
and since the fixed length of $\vec\alpha$ becomes an irrelevant
constraint in the limit $M\to\infty$,
each real degree of freedom
in $\alpha_q$ is an independent Gaussian
distributed number~\cite{mehta:90}. The squared modulus 
$|\alpha_q|^2$ has probability density
\begin{equation}
P(|\alpha_q|^2)= 
\frac{1}{2 \pi|\alpha_q|^2}
\left(\frac{2\pi^3|\alpha_q|^2}{w\Delta}\right)^{\beta/2}
\exp\left[-\frac{\beta\pi^2}{2 w \Delta}|\alpha_q|^2\right]
.
\label{eq:psq}
\end{equation}
Eq.\ (\ref{eq:psq})
is a $\chi^2$-distribution with $\beta$ degrees of freedom and mean
$\Delta w/\pi^2$.

We first determine the distribution of the
decay rate $\Gamma$ of the lasing mode,
following Ref.\ \cite{misirpashaev:98a}.
Since the lasing mode is the mode closest to the real axis, its decay
rate is much smaller than the typical decay rate of a mode, which is
$\simeq T\Delta$.
Then we calculate the conditional distribution
and mean of the Petermann factor for given $\Gamma$. The unconditional
distribution of the Petermann factor is found by folding
the conditional distribution with the distribution of $\Gamma$, but will
not be considered here.

\subsection{Decay rate of the lasing mode}

The amplification with rate $1/\tau_a$ is assumed to be effective
over a window $\Omega_a =L\Delta$
containing many modes. The lasing mode is the mode within this
window that has the
smallest decay rate $\Gamma$. For such small decay rates we can use
first-order perturbation theory to obtain  
the decay rate of mode $q$,
\begin{equation}
\Gamma_q=2\pi|\alpha_q|^2.
\end{equation}
The $\chi^2$-distribution (\ref{eq:psq})
of the squared moduli $|\alpha_q|^2$ translates into
a $\chi^2$-distribution of the decay rates,
\begin{equation}
P(\Gamma)\propto \Gamma^{(2-\beta)/2}\exp\left(
-\frac{\beta \pi\Gamma}{4w\Delta}\right)
.
\label{eq:pgammachi}
\end{equation}

Ignoring correlations, we may obtain
the decay rate of the lasing mode by
considering the $L$
decay rates as independent
random variables drawn from the distribution $P(\Gamma)$.
The
distribution of the smallest among the $L$ decay rates is then given by
\begin{equation}
P_L(\Gamma)=L P(\Gamma)\left[1-\int_0^\Gamma d\Gamma'\,P(\Gamma')\right]^{L-1}.
\label{eq:pl}
\end{equation}
For small rates $\Gamma$ we can insert the distribution (\ref{eq:pgammachi})
and obtain
\begin{mathletters}
\begin{eqnarray}
P_L(\Gamma)&\approx& \frac{1}{\sqrt{\Gamma}}\exp\left(-\frac{L\pi\Gamma}{4 w \Delta}
\right)\left[
\erf\left(\frac{\pi\Gamma}{4w\Delta} \right)\right]^{L-1}
,\ \beta=1,
\nonumber\\
\\
P_L(\Gamma)&\approx& \exp\left(-\frac{L\pi\Gamma}{2w\Delta}
\right),\quad \beta=2.
\end{eqnarray}
\end{mathletters}%
Here $\erf(x)=2\pi^{-1/2}\int_0^x dy\,\exp(-y^2)$ is the error function.
The decay rate of the lasing mode decreases with increasing width of
the amplification window as
$\Gamma \sim w\Delta (\Omega_a/\Delta)^{-2/\beta}\ll w\Delta $. 

\subsection{First-order perturbation theory}
\label{sec:firstorder}

If the opening is much smaller than a wavelength, then a perturbation theory in
$\vec{\alpha}$ seems a natural starting point.
We assign the index $l$ to the lasing mode, and 
write the perturbed right eigenfunction
$|l\rangle'=\sum_q d_q|q\rangle$ and the perturbed left
eigenfunction
$\langle l|'=\sum_q e_q \langle q|$,
in terms of the eigenfunctions of $H$.
The coefficients are $d_q=U_{ql}/U_{ll}$ and $e_q=U^{-1}_{lq}/U^{-1}_{ll}$,
i.\,e.,
we do not normalize the perturbed eigenfunctions but rather choose $d_l=e_l=1$.

To leading order the lasing mode remains at $\Omega=\omega_l$ and has width
\begin{equation}
\Gamma=2\pi |\alpha_l|^2
.
\end{equation}
The coefficients of the wave function are
\begin{equation}
d_q=i\frac{\pi \alpha_q\alpha_l^*}{\omega_q-\omega_l},
\qquad
e_q=i\frac{\pi \alpha_q^*\alpha_l}{\omega_q-\omega_l}.
\label{eq:wavecoef}
\end{equation}
The Petermann factor of the lasing mode follows from 
Eq.\ (\ref{eq:Kgl1}),
\begin{eqnarray}
K&=&\frac{\left(1+\sum_{q\neq l}|d_q|^2\right)
\left(1+\sum_{q\neq l}|e_q|^2\right)}{\left|1+\sum_{q\neq l}d_q e_q \right|^2}
\nonumber
\\
&\approx&1+\sum_{q\neq l}|d_q-e_q^*|^2
,
\label{eq:Kgl1a}
\end{eqnarray}
where we linearized with respect to $\Gamma$ because the lasing mode is
close to the real axis.
From Eq.\ (\ref{eq:wavecoef})
one finds
\begin{equation}
	K = 1 + (2 \pi |\alpha_l|)^2 \sum_{q\ne l}
	\frac{|\alpha_q|^2}{(\omega_l-\omega_q)^2} .
	\label{eq:Kgl2}
\end{equation}
We seek the distribution $P(K)$ and the
average $\langle K \rangle_{\Omega,\Gamma}$ of
$K$ for a given value of $\Omega$ and $\Gamma$.

For $\beta=1$,
the probability to find an eigenvalue at
$\omega_q$ given that there is an eigenvalue at $\omega_l$ vanishes {\em
linearly} for small $|\omega_q-\omega_l|$, as a consequence of eigenvalue
repulsion constrained by time-reversal symmetry. Since the expression
(\ref{eq:Kgl2}) for $K$ diverges {\em quadratically} for small
$|\omega_q-\omega_l|$, we conclude
that $\langle K \rangle_{\Omega,\Gamma}$ does
not exist in perturbation theory%
\footnote{
For broken time-reversal symmetry
there is no divergence. We can use the known
two-point correlation
function $R(\omega_l,\omega_q)$ of the Gaussian unitary ensemble
to obtain
$\langle K \rangle_{\Omega,\Gamma}=1+\frac{1}{3}\pi T\Gamma /\Delta $
for $T\ll 1$.}.
This severely complicates the
problem.

\subsection{Summation of the perturbation series}

To obtain a finite answer for the average Petermann factor
we need to go beyond perturbation theory. By a complete summation of the
perturbation series we will in this section obtain results that 
are valid for all values $T\le 1$ of the transmission probability.
Our starting point are the exact relations
\begin{mathletters}
\begin{eqnarray}
	d_q z_l &=& \omega_q d_q - i \pi \alpha_q \sum_{p} \alpha_p^* d_p,\\
	e_q z_l &=& \omega_q e_q - i \pi \alpha_q^* \sum_{p} \alpha_p e_p,
\end{eqnarray}
\end{mathletters}%
between the complex eigenvalues $z_q$ of ${\cal H}$
and the real eigenvalues $\omega_q$
of $H$. Distinguishing between $q=l$ and $q\ne l$,
we obtain three recursion relations,
\begin{mathletters}
\begin{eqnarray}
	z_l &=& \omega_l - i \pi |\alpha_l|^2
		- i \pi \alpha_l \sum_{q\ne l} \alpha_q^* d_q,	\\
	i d_q &=& \frac{\pi \alpha_q}{z_l - \omega_q} \biggl( \alpha_l^* 
		+ \sum_{p\ne l} \alpha_p^* d_p \biggr), \\
	i e_q &=& \frac{\pi \alpha_q^*}{z_l - \omega_q} \biggl( \alpha_l 
		+ \sum_{p\ne l} \alpha_p e_p \biggr).
	\label{tempeq13b}
\end{eqnarray}
\end{mathletters}

We now use the fact that $z_l$ is the eigenvalue closest to the real axis. We
may therefore assume that $z_l$ is close to the unperturbed value $\omega_l$
and replace the denominator $z_l-\omega_q$ in Eq.\ (\ref{tempeq13b}) by
$\omega_l-\omega_q$. That decouples the recursion
relations, which may then be solved
in closed form,
\begin{mathletters}
\begin{eqnarray}
	z_l &=& \omega_l - i \pi |\alpha_l|^2 \left( 1 + i \pi A \right)^{-1}
		,\\
	i d_q &=& \frac{\pi \alpha_q \alpha_l^*}{\omega_l-\omega_q}
		\left(1 + i \pi A \right)^{-1} , \\
	i e_q &=& \frac{\pi \alpha_q^* \alpha_l}{\omega_l-\omega_q}
		\left(1 + i \pi A \right)^{-1} .
\end{eqnarray}
\end{mathletters}%
We have defined 
\begin{equation}
A=\sum_{q\ne l} |\alpha_q|^2 (\omega_l-\omega_q)^{-1}
.
\label{eq:a}
\end{equation}
The decay
rate of the lasing mode is
\begin{equation}
\Gamma = - 2 \,\imagteil z_l = 2 \pi |\alpha_l|^2 (1+\pi^2 A^2)^{-1}.
\label{eq21}
\end{equation}
From Eq.\ (\ref{eq:Kgl1a}) we find
\begin{equation}
K =  1 + \frac{2\pi\Gamma}{\Delta}\frac{B}{1+\pi^2 A^2} ,
\label{tempeq23}
\end{equation}
with 
\begin{equation}
B=\Delta\sum_{q\ne l} |\alpha_q|^2 (\omega_l - \omega_q )^{-2}
.
\label{eq:b}
\end{equation}

The problem is now reduced to a calculation
of the joint probability distribution
$P(A,B)$. This problem is closely related to the
level curvature problem of
random-matrix theory~\cite{oppen:9x,fyodorov:95a}.
The calculation is presented in Appendix \ref{app:a}. The result is
\begin{eqnarray}
P(A,B)&=&
\frac{\pi}{24} \left(
\frac{8}{\pi w}\right)^{\beta/2}
\frac{(\pi^2A^2+w^2)^\beta}{B^{2 +3\beta/2}}
\nonumber
\\
&&{}\times
\exp\left[-\frac{\beta w}{2 B}\left(\frac{\pi^2
A^2}{w^2}+1\right)
\right]
.
\label{eq:pab}
\end{eqnarray}

\subsection{Probability distribution of the Petermann factor}

From
Eqs.\ (\ref{eq:psq}), (\ref{eq21}), (\ref{tempeq23}),
and (\ref{eq:pab}) we can
compute the probability distribution
\begin{mathletters}
\begin{eqnarray}
P(K)&=&
\left\langle Z \right\rangle^{-1}
\left\langle
\delta\left(K-1-\frac{2\pi\Gamma}{\Delta}\frac{B}{1+\pi^2A^2}\right)Z
\right\rangle
,
\nonumber
\\
\\
Z & = & \delta(\Omega-\omega_l) \delta\left(
\Gamma - \frac{2 \pi |\alpha_l|^2}{ 1 + \pi^2 A^2} \right) 
,
\label{eq:z}
\end{eqnarray}
\end{mathletters}%
of $K$ at fixed $\Gamma$ and $\Omega$
by averaging over $|\alpha_l|^2$, $A$, and $B$.
In principle one should also require that the decay rates of modes $q\ne l$ are
bigger than $\Gamma$, but this extra condition becomes irrelevant for
$\Gamma\to 0$.
The average of $Z$ over $|\alpha_l|^2$ with Eq.\ (\ref{eq:psq})
yields a factor $(1+\pi^2 A^2)^{\beta/2}$. (Only the behavior of
$P(|\alpha_l|^2)$ for small $|\alpha_l|^2$ matters, because we concentrate on
the lasing mode.)
After integration over $B$
the distribution
can be expressed as a ratio of integrals
over $A$,
\begin{eqnarray}
&&P(K)=\frac{(2\pi)^{2\beta}}{3\beta}
\frac{\Delta w}{\Gamma}\left(\frac{(K-1)\Delta}{w\Gamma}\right)^{-2-3\beta/2}
\nonumber\\
&&{}\times
\int_0^\infty \!\!\! dA\,\frac{(1+\pi^2A^2/w^2)^\beta}{(1+\pi^2A^2)^{1+\beta}}
\exp\left[-\frac{\beta\pi w \Gamma(1+\pi^2A^2/w^2)}{(K-1)\Delta
(1+\pi^2A^2)}
\right]
\nonumber\\
&&{}\times
\left(
\int_0^\infty dA\,\frac{(1+\pi^2A^2)^{\beta/2}}{(1+\pi^2A^2/w^2)^{1+\beta/2}}
\right)^{-1}
.
\label{eq:kdist}
\end{eqnarray}

\begin{figure}
\epsfig{file=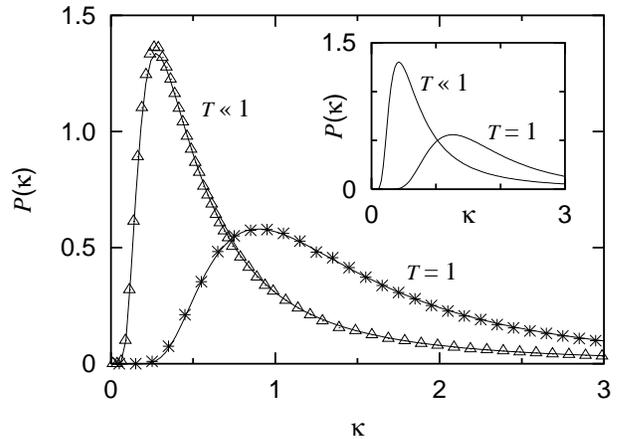,width=3.4in}
\medskip
\caption{Probability distribution of the rescaled Petermann factor
$\kappa = (K - 1)\Delta / \Gamma T$
for $T=1$ and $T\ll 1$, in the presence of time-reversal symmetry.
The solid curves follow from
Eqs.\ (\ref{distfunk}) (with $\beta=1$) and (\ref{tempeq19}).
The data points follow from a numerical simulation of the random-matrix model.
The inset shows the results (\ref{distfunk}) (with $\beta=2$)
and (\ref{tempeq19a})
for broken time-reversal symmetry.
}
\label{figeinsklein}
\end{figure}

We introduce the rescaled Petermann factor
$\kappa=(K-1)\Delta/\Gamma T$.
A simple result for $P(\kappa)$ follows for $T=1$,
\begin{equation}
	P(\kappa)=
\frac{4\beta  \pi^{2\beta}}{3\kappa^{2+3\beta/2} }
\exp\left[-\frac{\beta \pi}{\kappa}\right] ,
	\label{distfunk}
\end{equation}
and for $T\ll 1$, 
\begin{mathletters}
\begin{equation}
	P(\kappa)=\frac{\pi}{12 \kappa^2}\left(1+\frac{\pi}{2\kappa}\right)
	\exp\left[-\frac{\pi}{4\kappa}\right]
	, \quad \beta=1,
\label{tempeq19}
\end{equation}
\begin{equation}
P(\kappa)=\frac{\pi}{ 8\sqrt{2\kappa^5}}
\left(1+\frac{2\pi}{3\kappa}+\frac{\pi^2}{3\kappa^2}
\right)
\exp\left[-\frac{\pi}{2\kappa}\right]
, \quad \beta=2 .
\label{tempeq19a}
\end{equation}
\end{mathletters}%
As shown in Fig.\ \ref{figeinsklein}, the distributions are
very broad and
asymmetric, with a long tail towards large
$\kappa$.

To check
our analytical results
we have also done a numerical simulation of the random-matrix model,
generating a large number of random matrices $H$ and computing
$K$ from Eq.\ (\ref{eq:Kgl1}). As one can see from Fig.\ \ref{figeinsklein},
the agreement with the theoretical predictions is flawless.

\subsection{Mean Petermann factor}

\begin{figure}
\epsfig{file=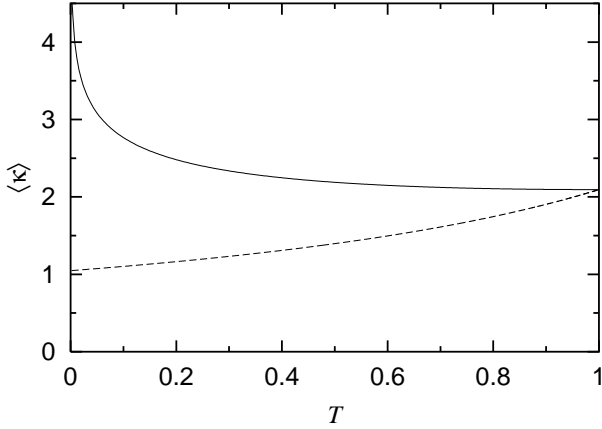,width=3.4in}
\medskip
\caption{Average of the rescaled Petermann factor $\kappa$
as a function of 
transmission probability $T$.
The solid curve is the result~(\ref{tempeq22})
in the presence of time-reversal symmetry,
the dashed curve is the result~(\ref{eq24}) for broken time-reversal symmetry.
For small $T$, the solid curve diverges $\propto \ln T^{-1}$ while the dashed
curve has the finite limit of $\pi/3$. For $T=1$ both curves reach the value
$2\pi/3$.
}
\label{peter1fig}
\end{figure}

The distribution (\ref{eq:kdist}) 
gives 
for preserved time-reversal symmetry ($\beta=1$) the mean Petermann factor
\begin{equation}
\left\langle K\right\rangle_{\Omega,\Gamma}=1-\frac{\Gamma}{\Delta}
\frac{2\pi}{3} \frac{G^{22}_{22}\left(
w^2\left|\begin{array}{cc}0 & 0\\ -\case{1}{2} & -\case{1}{2}
\end{array}\right.\right)
}{
G^{22}_{22}\left(
w^{2}\left|\begin{array}{cc}
-\case 12 &\case 12\\-1 & 0
\end{array}\right.\right)}
,
\label{tempeq22}
\end{equation}
in terms of the ratio of two Meijer $G$-functions.
We have plotted the result in Fig.\ \ref{peter1fig}, as a function of $T=4
w(1+w)^{-2}$.

It is remarkable that the average $K$ depends
{\em non-analytically} on $T$, and hence on the area of
the opening.
(The transmission probability $T$ is related to the area
${\cal A}$ of the opening by $T \simeq {\cal A}^3/
\lambda^6$ for $T\ll 1$ \cite{Bethe}.)
For $T\ll 1$, the average approaches the form
\begin{equation}
	\langle K \rangle_{\Omega,\Gamma} = 1 + \frac{\pi}{6}
	\frac{T \Gamma}{\Delta} \ln \frac{16}{T} .
	\label{tempeq21}
\end{equation}
The most
probable (or modal) value of $K-1\simeq T\Gamma/\Delta$ is
parametrically
smaller than the mean value (\ref{tempeq21}) for $T\ll 1$.
The non-analyticity results from the
relatively weak eigenvalue repulsion in the presence of time-reversal symmetry.
If time-reversal symmetry is broken,
then the stronger quadratic repulsion is
sufficient to overcome the $\omega^{-2}$ divergence of 
perturbation theory (\ref{eq:Kgl2}) and
the average $K$ becomes an analytic function of $T$. For this case, we find
from Eq.\ (\ref{eq:kdist})
the mean Petermann factor
\begin{equation}
	\langle K \rangle_{\Omega,\Gamma}
	= 1 + \frac{\Gamma}{\Delta} \frac{4 \pi w}{3
	(1+w^2)} ,
	\label{eq24}
\end{equation}
shown dashed in Fig.\ \ref{peter1fig}.

\section{Many scattering channels}
\label{sec:manychannels}

For arbitrary number of scattering channels $N$ the coupling 
matrix $W$ is an $M\times N$ rectangular matrix. The square matrix
$\pi W^\dagger W$ has $N$ eigenvalues $(M\Delta/\pi)w_n$.
The transmission coefficients
of the eigenchannels are 
\begin{equation}
T_n=\frac{4w_n}{(1+w_n)^2}.
\end{equation}
A single hole of area 
${\cal A}\gg \lambda^2$ (at wavelength $\lambda$)
corresponds to $N\simeq 2\pi {\cal A}/\lambda^2$
fully transmitted scattering channels, with all $T_n=w_n=1$ the same.

As in the single-channel case, we first determine the distribution of
the decay rate $\Gamma$ of the lasing mode. This decay rate
is smaller than the
typical decay rate $\Gamma_0=TN\Delta/2\pi$ of the non-lasing modes.
Then we calculate the mean Petermann factor $\langle
K\rangle$ for given $\Gamma$
and investigate its behavior for the atypically small
decay rates of the lasing mode.

\subsection{Decay rate of the lasing mode}
\label{sec:lthresh}

\begin{figure}
\epsfig{file=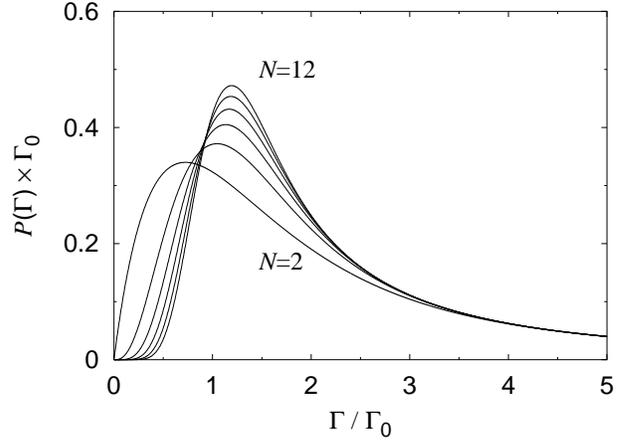,width=3.4in}
 \medskip
\caption{
Decay-rate
distribution $P(\Gamma)$ of a chaotic cavity with an opening 
that supports $N=2$, $4$, $6$, $8$, $10$, $12$
fully transmitted scattering channels. Computed from Eq.\
(\protect\ref{eq:pgue}), for the case of broken time-reversal
symmetry.
}
\label{fig:pgamma}
\end{figure}

The distribution of decay rates $P(\Gamma)$
has been calculated
by Fyodorov and Sommers. For broken time-reversal symmetry
the result is \cite{fyodorov:96a,fyodorov:97a}
\begin{mathletters}
\begin{eqnarray}
P(\Gamma)&=&\frac{\pi}{\Delta}
{\cal F}_1\left(\frac{\pi}{\Delta}\Gamma\right)
{\cal F}_2\left(\frac{\pi}{\Delta}\Gamma\right)
,
\\
{\cal F}_1(y)&=&
\frac{1}{2\pi}\int_{-\infty}^\infty dx\, e^{-ixy}
\prod_{n=1}^N\frac{1}{g_n-ix},
\label{eq:calfg1}
\\
{\cal F}_2(y)&=&
\frac{1}{2}\int_{-1}^1 dx\, e^{-xy}\prod_{n=1}^N(g_n+x),
\label{eq:calfg2}
\end{eqnarray}
\label{eq:pgue}%
\end{mathletters}%
where $g_n=-1+2/T_n$.
For identical $g_n\equiv g$ the two functions ${\cal F}_1$ and ${\cal
F}_2$ simplify to
\begin{mathletters}
\begin{eqnarray}
{\cal F}_1(y) &=&
\frac{1}{(N-1)!}y^{N-1}
e^{-gy}
,
\\
{\cal F}_2(y)&=&
\sum_{n=0}^N(-1)^n{N\choose n}g^{N-n}
\frac{d^n}{dy^n}\left(\frac{\sinh y}{y}\right)
,
\end{eqnarray}
\label{eqs:calf}%
\end{mathletters}
and a convenient form of the distribution function is
\begin{equation}
P(\Gamma)=\frac{\Delta}{2\pi\Gamma^2 (N-1)!}
\int_{N(1-T)\Gamma/\Gamma_0}^{N\Gamma/\Gamma_0}dx\,x^Ne^{-x}
.
\label{eq:pgcon}
\end{equation}
The behavior of $P(\Gamma)$ for various numbers $N$ of
fully transmitted ($T=1$) scattering
channels is illustrated in Fig.\ \ref{fig:pgamma}.

The result for preserved time-reversal symmetry is a bit more
involved \cite{fyodorov:99a}. Fortunately, we can draw all important
conclusions from the results for broken time-reversal
symmetry, on which we will concentrate here.

For large $N$, the
distribution $P(\Gamma)$ becomes non-zero only in the interval
$\Gamma_0<\Gamma<\Gamma_0/(1-T)$, where it is equal to
\cite{haake:92,lehmann:95}
\begin{equation}
P(\Gamma)=\frac{\Gamma_0}{T\Gamma^2}
,\quad \Gamma_0<\Gamma<\Gamma_0/(1-T).
\label{eq:pgammalargen}
\end{equation}
This limit is $\beta$-independent.
The smallest decay rate $\Gamma_0$ corresponds to the inverse mean dwell time
in the cavity.

We are interested in the ``good cavity'' regime, where the typical decay rate
$\Gamma_0$ is small compared to the
amplification bandwidth $\Omega_a$. 
From $\Gamma_0=TN\Delta/2\pi$ it follows  that
the number $L\simeq\Omega_a/\Delta$ of
amplified modes is then much larger than $TN$.
In this regime the decay rate of the lasing mode
(the smallest among the $L$ decay
rates in the frequency window $\Omega_a$) drops below $\Gamma_0$.
The asymptotic result (\ref{eq:pgammalargen}) cannot be used in this
case,
since it does not describe accurately the tail $\Gamma\lesssim
\Gamma_0$. Going back to the exact result (\ref{eq:pgue}) 
we find for the tail of the distribution the expression
\begin{equation}
P(\Gamma)=\frac{\pi}{N T^2 \Delta}[1+\erf(u)]+{\cal O}(N^{-3/2}),
\label{eq:pu}
\end{equation}
where we have defined
$u=\sqrt{N/2}  (\Gamma/\Gamma_0-1)$.

The distribution
$P_L(\Gamma)$ of the lasing mode follows from $P(\Gamma)$ by means of
Eq.\ (\ref{eq:pl}). We find that it
has a pronounced maximum
at a value $u_{\rm max}$ determined by 
\begin{equation}
\frac{\exp(-u^2_{\rm max})}{[1+\erf(u_{\rm max})]^2}=\frac{L-1}{\sqrt{2N}}
\frac{\sqrt{\pi}(g+1)}{4}.
\end{equation}
For $L\gg \sqrt{N}$ (and hence also in the good cavity regime) we find 
$u_{\rm max}\sim -\sqrt{\ln L}<0$, and the deviation of $\Gamma$ from
$\Gamma_0$
is of order $\Delta \sqrt{N}\ll \Gamma_0$ (as long as $L\ll e^N$).

\subsection{Mean Petermann factor}
\label{sec:results}

Eigenfunction correlations of non-Hermitian operators have been studied 
in Refs.\ \cite{janik:99a,chalker:98a}. The eigenfunction autocorrelator
considered in these studies is directly connected to the Petermann
factor $K$. Ref.\ \cite{janik:99a} provides a convenient
expression of the mean Petermann factor,
\begin{eqnarray}
&&M \pi \langle K \rangle_{\Omega,\Gamma}\rho(\omega)\nonumber\\
&&=\lim_{\veps\to 0^+}\left\langle
\left(
\spur\frac{\varepsilon}{(\omega-\hmat)(\omega^*-\hmat^\dagger)+\varepsilon^2}
\right)^2
\right\rangle
.
\label{eq:genf}
\end{eqnarray}
In Ref.\ \cite{janik:99a}
this average has been calculated perturbatively for $N\gg 1$,
with the result 
\begin{equation}
\langle K\rangle_{\Omega,\Gamma}\approx
-N\left(\frac{\Gamma}{\Gamma_0}-1\right)
\left(\frac{(1-T)\Gamma}{\Gamma_0}-1\right)
\label{eq:meankpert1}
\end{equation}
for $\Gamma_0<\Gamma<\Gamma_0/(1-T)$.
This result is at the same level of approximation
as Eq.\ (\ref{eq:pgammalargen}) for the distribution of the decay rates,
i.\,e.\ it does not describe the range $\Gamma\lesssim\Gamma_0$ of
atypically small decay rates. Since that is precisely the range that
we need for the Petermann factor, we can not use the existing
perturbative results.
We have
calculated the mean Petermann factor non-perturbatively for any $\Gamma$
and $N$, assuming broken time-reversal symmetry.
The derivation is given in Appendix \ref{app:b}.
The final result for the mean Petermann factor is
\begin{mathletters}
\begin{eqnarray}
\langle K\rangle_{\Omega,\Gamma}&=&1+\frac{2S(\pi\Gamma/\Delta)
}{{\cal F}_1(\pi\Gamma/\Delta){\cal F}_2(\pi\Gamma/\Delta)}
,
\\
S(y)&=& - \int_0^ydy'\, {\cal F}_1(y')\frac{\partial}{\partial y'}
{\cal F}_2(y')
,
\end{eqnarray}
\label{eqs:meankfinal}%
\end{mathletters}
with ${\cal F}_1$ and ${\cal F}_2$ given in Eq.\ (\ref{eq:pgue}).

For identical
$g_n\equiv g$ we can use Eq.\ (\ref{eqs:calf}) and obtain
by successive integrations by parts
\begin{equation}
S(y)=
\sum_{n=0}^{N-1}\frac{(-1)^n}{n!}y^n
\frac{d^n}{dy^n}
\left\{
e^{-gy}\frac{d}{dy}
\left(
\frac{\sinh y}{y}
\right)
\right\}
.
\end{equation}
For $N=1$ and $\Gamma\ll \Delta$ we recover the single-channel
result (\ref{eq24}) of the previous section.
In what follows we will continue to assume for simplicity that all
$g_n$'s are equal to a common value $g$.

The large-$N$ behavior can be conveniently
studied from the expression
\begin{eqnarray}
S(y)&=&-\frac{1}{4y^2(N-1)!}
\int_{y(g-1)}^{y(g+1)}
dx\,
x^{N-1}e^{-x}
\nonumber\\
&&{} \times
[x-(g-1)y] [x-(g+1)y]
,
\label{eq:scon}
\end{eqnarray}
because the integral permits a saddle-point approximation.
For $\Gamma>\Gamma_0$ we
recover Eq.\ (\ref{eq:meankpert1}), but now we can also study the
precise 
behavior of the mean Petermann factor for $\Gamma\lesssim \Gamma_0$,
hence also for decay rates relevant for the lasing mode.
The results will again be presented in terms of the rescaled
parameter $u=\sqrt{N/2}(\Gamma/\Gamma_0-1)$.
We expand the integrands in Eqs.\ (\ref{eq:pgcon}) and
(\ref{eq:scon}) around the
saddle point at $x=N$ (which coincides with the upper integration
limit at $\Gamma=\Gamma_0$)  
and keep the first non-Gaussian correction. This yields
\begin{mathletters}
\begin{eqnarray}
\langle
K\rangle_{\Omega,\Gamma}&=&
T\sqrt{2N}[F(u)+u] -T(g-1)\,u^2
\nonumber
\\ &&{}
+
T F(u) \left[(3-g)u+\case{4}{3}u^3+ \case{4}{3}(1+u^2)\,F(u)\right]
\nonumber
\\ &&{}
+
{\cal O}(N^{-1/2}),
\\
F(u)&=& \frac{\exp(-u^2)}{\sqrt{\pi}
\,[1+\mbox{erf} (u)]}
.
\end{eqnarray}
\end{mathletters}
For $\Gamma=\Gamma_0$ ($u=0$) this simplifies to
\begin{equation}
\langle K\rangle_{\Omega,\Gamma=\Gamma_0}
=T\left(\sqrt{\frac{2N}{\pi}}+\frac{4}{3\pi} \right).
\label{eq:ksqrtn}
\end{equation}

\begin{figure}
\epsfig{file=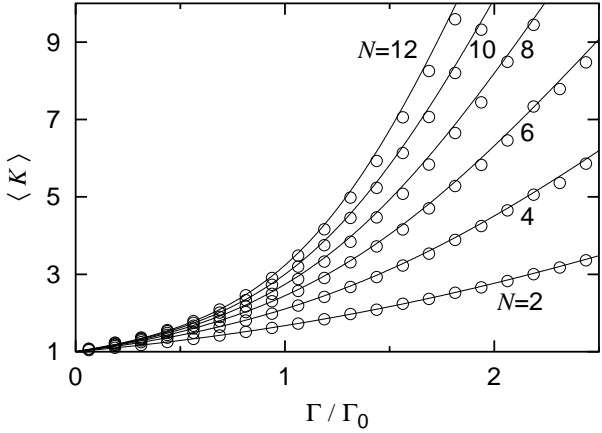,width=3.4in}
 \medskip
 \caption{Average Petermann factor $\langle K\rangle$ as a function of the
decay rate $\Gamma$ for different values $N$ of fully
transmitted scattering channels.
The solid curves are the analytical result~(\ref{eqs:meankfinal}),
the data points are obtained by
a numerical simulation. Time-reversal symmetry is
broken.
}
\label{fig:mean1}
\end{figure}

We see that the
mean Petermann factor varies on the same scale of $\Gamma$ as the 
decay-rate distribution $P(\Gamma)$, Eq.\ (\ref{eq:pu}). However, while
$P(\Gamma)$ decays exponentially for $u\ll-1$, the mean Petermann factor
displays an algebraic tail 
\begin{equation}
\langle K\rangle_{\Omega,\Gamma} =-\frac{T\sqrt{N}}{u\sqrt{2}}+1-T+{\cal
O}(u^{-2}).
\end{equation}
For an amplification window $\Omega_a=L\Delta$ with $L\gg
\sqrt{N}$
we found in Section \ref{sec:lthresh}
that the decay rate $\Gamma$  of the lasing mode drops below $\Gamma_0$
(the rescaled parameter $u_{\rm max}\sim -\sqrt{\ln L}$).
Still, the mean Petermann factor
\begin{equation}
\langle K\rangle_{\Omega,\Gamma}
\sim \sqrt{\frac{N}{\ln L}}
\end{equation}
remains parametrically larger than unity (as long as $L\ll \sqrt{N}e^N$).

\begin{figure}
\epsfig{file=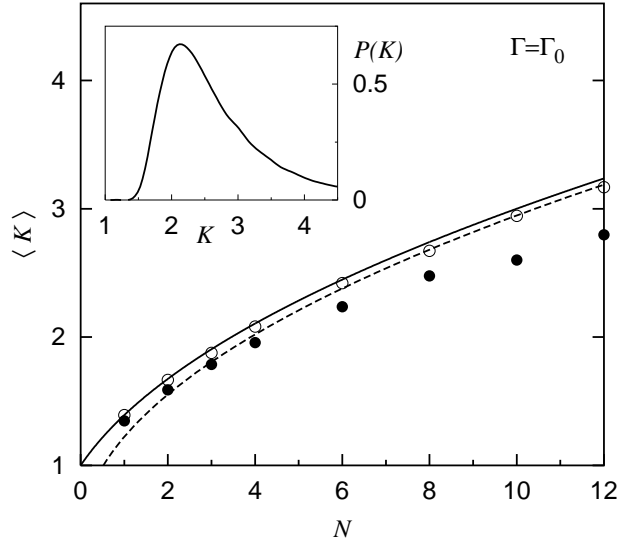,width=3.4in}
 \medskip
\caption{Average of the Petermann factor $K$ at $\Gamma=\Gamma_0$ as
function of the number $N$
of fully transmitted scattering channels. The analytical
result (\ref{eqs:meankfinal}) for broken time-reversal symmetry (full curve)
is compared with the result of a numerical
simulation (open circles for broken time-reversal symmetry,
filled circles for preserved time-reversal symmetry).
The dashed line is the large-$N$ result (\ref{eq:ksqrtn}).
The inset shows the distribution of 
$K$ at $\Gamma=\Gamma_0$ for $N=10$.
}
\label{fig:ksqrtn}
\end{figure}

We now compare
our analytical findings with the results of numerical simulations.
We generated a large number of random matrices ${\cal H}$ with
dimension $M=120$ ($M=200$) for $N=2,4,6,8$ ($N=10,12$)
fully transmitted scattering channels ($g=T=1$).
Fig.\ \ref{fig:mean1} shows the mean $K$ at given $\Gamma$.
We find excellent
agreement with our analytical result (\ref{eqs:meankfinal}).

The behavior $\langle K\rangle\sim\sqrt{N}$ at
$\Gamma=\Gamma_0$ is shown in Fig.\ \ref{fig:ksqrtn}.
The inset depicts the
distribution of $K$ at $\Gamma=\Gamma_0$ for $N=10$,
which only can be accessed numerically. We see that the mean Petermann
factor is somewhat larger than the most probable (or modal) value.

\subsection{Preserved time-reversal symmetry}
\label{sec:preserved}

\begin{figure}
\epsfig{file=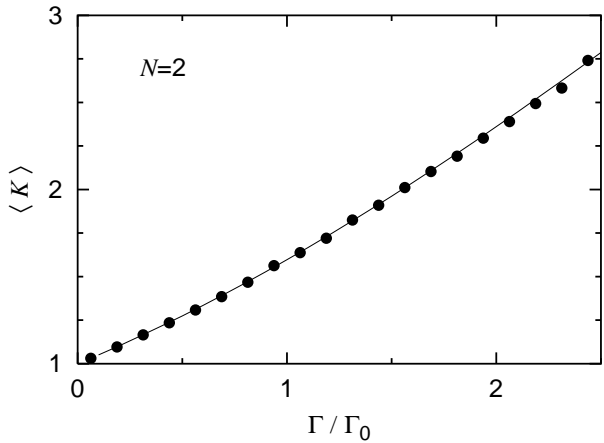,width=3.4in}
 \medskip
\caption{Theoretical expectation (\protect\ref{eq:goen2})
(full curve) and
the result of a numerical simulation (data points)
for the average Petermann factor $K$ 
in the presence of time-reversal symmetry,
as a function of the
decay rate $\Gamma$ for 
$2$ fully transmitted scattering channels.
}
\label{fig:goen2}
\end{figure}

In the derivation of the
mean Petermann factor for broken time-reversal symmetry
(Appendix \ref{app:b})
it turned out that the final result is formally
connected to the  expression for the decay-rate 
distribution $P(\Gamma)$, inasmuch as  both expressions are built from the
factors ${\cal F}_1$ (involving non-compact bosonic degrees of
freedom of the saddle-point manifold)
and ${\cal F}_2$ (involving compact bosonic degrees of
freedom of that manifold). We tried to translate this
description to the case of preserved time-reversal symmetry ($\beta=1$), by
operating in the same way on the compact and non-compact factors of the
expression of Ref.\ \cite{fyodorov:99a},
but could obtain a satisfactory result only for
$N=2$, 
\begin{equation}
\langle K\rangle=\frac{1}{2\Gamma_0}\frac{\Gamma(\Gamma-\Gamma_0)
\exp(\Gamma/\Gamma_0)
+\Gamma_0^2\sinh(\Gamma/\Gamma_0)}{
\Gamma\cosh (\Gamma/\Gamma_0)  - \Gamma_0\sinh (\Gamma/\Gamma_0)}.
\label{eq:goen2}
\end{equation}
In Fig.\ \ref{fig:goen2}
this expression is compared to the result of a numerical simulation.

\begin{figure}
\epsfig{file=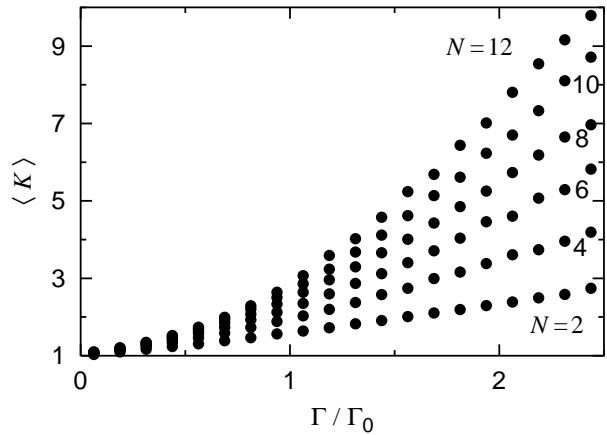,width=3.4in}
 \medskip
\caption{Results of a numerical simulation of the
average Petermann factor $\langle K\rangle$ 
in the presence of time-reversal symmetry,
as a function of the
decay rate $\Gamma$ for 
$N$ fully transmitted scattering channels.
}
\label{fig:mean2}
\end{figure}

\begin{figure}
\epsfig{file=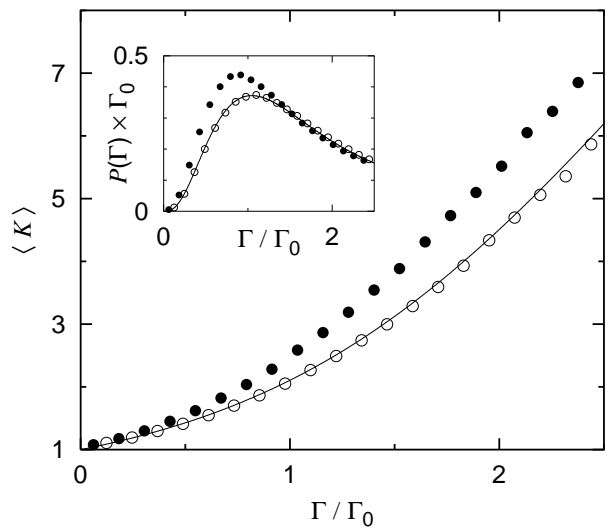,width=3.4in}
 \medskip
\caption{
Average Petermann factor $\langle K\rangle$ for $N=4$,
$\beta=2$ [open circles: result of a numerical simulation,
curve: Eq.\ (\ref{eqs:meankfinal})] and for $N=8$, $\beta=1$ (filled
circles: result of a numerical simulation).
The parameter $\Gamma_0$ equals $N\Delta/2\pi$ in both cases, so it is
twice as large for $\beta=2$ as for $\beta=1$.
The
inset depicts the probability distribution of $\Gamma$. 
}
\label{fig:fig6}
\end{figure}
For larger numbers of channels we can draw our conclusions from
the numerical results that are presented in Fig.\ \ref{fig:mean2}.
Interestingly enough the data points for $N$ channels
are close to the
results for broken time-reversal symmetry with $N/2$ channels,
when the decay rate
is given in units of $\Gamma_0$. This is illustrated for $N=8$
in Fig.\ \ref{fig:fig6}.
Such a rule of thumb (motivated by the number of real degrees of freedom
that enter the non-Hermitian part of ${\cal H}$) 
was already known for the decay rate distribution
(inset in Fig.\ \ref{fig:fig6}).
Hence the Petermann factor for the lasing mode should again display a
sublinear growth with increasing channel number $N$. This expectation is
indeed
confirmed by the numerical simulations, see the filled circles in
Fig.\ \ref{fig:ksqrtn}.

\section{Discussion}
\label{sec:conclusions}

The Petermann factor $K$ enters the fundamental lower limit of the laser
linewidth due to vacuum fluctuations and is a measure of the
non-orthogonality of cavity modes.
We related the Petermann factor to the residue of the scattering-matrix
pole that pertains to the lasing mode and
computed statistical properties of $K$
in an ensemble of chaotic cavities.
The technical
complications that had to be overcome arise from the fact
that laser action selects a mode which has a small decay rate
$\Gamma$, and hence belongs to
a pole that lies anomalously close to the real axis.
Parametrically large Petermann factors $\propto \sqrt N$
arise when the number $N$ of scattering
channels is large. For a single scattering channel the mean Petermann factor
depends non-analytically on the transmission probability $T$.

The quantity $K$ is also of fundamental significance in the general
theory of scattering resonances, where it enters the width-to-height
relation of resonance peaks and
determines the
scattering strength of a quasi-bound state with given
decay rate $\Gamma$. If we write the scattering matrix
(\ref{Smatrix}) in the form
\begin{equation}
S_{nm}=\delta_{nm}+\sigma_n^{}\sigma_m'(\omega-\Omega+i\Gamma/2)^{-1}
,
\end{equation}
then the scattering strengths $\sigma_n^{}$, $\sigma_m'$
are related to $\Gamma$
by a sum rule. For
resonances close to the real axis
($\Gamma\ll\Delta$) the relation is
\begin{equation}
\label{eq:sumrule}
\sum_{n,m}|\sigma_n^{}\sigma_m'|^2=\Gamma^2
\ .
\end{equation}
For poles deeper in the complex plane, however, the sum rule has to be
replaced by
\begin{equation}
\label{eq:sr2}
\sum_{n,m}|\sigma_n^{}\sigma_m'|^2=K\Gamma^2\ ,\quad K\geq 1\ .
\end{equation}
The method of filter diagonalization (or harmonic
inversion) that was used in Ref.\ \cite{harminv} to obtain
for the $H_3^+$ molecular ion
the location
of poles even deep in the complex plane
can also be employed
to determine the corresponding residues, and hence $K$.

The parameter $K$ defined in Eq.\ (\ref{eq:Kgl1}) appears as a
measure of mode non-orthogonality
also in problems outside of scattering theory. These problems
involve non-Hermitian operators that are
not of the form (\ref{eq:hmat}) \cite{chalker:98a}.
Many applications share the
common feature that they can be addressed statistically by an ensemble
description, and that
the physically relevant modes lie at the
boundary of the complex eigenvalue spectrum.
The non-perturbative statistical
methods reported in this paper should prove useful
in the investigation of some of these problems as well.

\acknowledgements

We have benefitted from discussions with P.\ W.\ Brouwer,
Y.\ V.\ Fyodorov, and F.\ von Oppen.
This work was supported by the Nederlandse organisatie voor
Wetenschappelijk Onderzoek (NWO), the Stichting voor Fundamenteel
Onderzoek der Materie (FOM), and
by the
European Commission via the 
Program for the Training and Mobility of Researchers (TMR).

\appendix

\section{Joint distribution of $A$ and $B$}
\label{app:a}
We 
calculate the joint distribution $P(A,B)$ [Eq.\ (\ref{eq:pab})]
of the quantities $A$ [Eq.\
(\ref{eq:a})] and $B$ [Eq.\
(\ref{eq:b})] 
by generalizing the theory of Ref.~\cite{fyodorov:95a}.
We give the lasing mode $\omega_l$ the new index $M$ and assume that it lies at
the center of the  semicircle (\ref{eq:semicircle}), 
$\omega_M=0$. Other choices just renormalize the mean modal spacing $\Delta$,
which we can set to $\Delta=1$.
The quantities $A$ and $B$ are then of the form
\begin{equation}
A=
\sum_{m=1}^{M-1}\frac{|\alpha_m|^2}{\omega_m}
,\qquad
B=
\sum_{m=1}^{M-1}\frac{|\alpha_m|^2}{\omega_m^2}
.
\end{equation}
The joint probability distribution of $A$ and $B$,
\begin{eqnarray}
P(A,B)&=&
\left\langle
\delta\left(A-
\sum_{m=1}^{M-1}\frac{|\alpha_m|^2}{\omega_m}
\right)
\right.
\nonumber
\\
&&\left.{}\times
\delta\left(B-
\sum_{m=1}^{M-1}\frac{|\alpha_m|^2}{\omega_m^2}
\right)
\right\rangle
,
\label{eq:pabdef}
\end{eqnarray}
is obtained by averaging
over the variables $\{|\alpha_m|^2,\omega_m\}$.
The quantities $|\alpha_m|^2$ are independent numbers with probability
distribution (\ref{eq:psq}).
The joint probability distribution of the eigenfrequencies $\{\omega_m\}$
of the closed cavity
is the eigenvalue distribution
of the Gaussian
ensembles  (\ref{eq:hdist}) of random-matrix theory,
\begin{equation}
P(\{\omega_m\})\propto\prod_{i<j}|\omega_i-\omega_j|^\beta
\exp\left[-\frac{\beta M}{4\mu^2}\sum_k \omega_k^2\right].
\label{eq:jevaldist}
\end{equation}
Our choice $\Delta=1$ translates into $\mu=M/\pi$.

The joint probability distribution
of the eigenvalues $\{\omega_m\}$ ($m=1\ldots M-1$)
is found by setting $\omega_M=0$
in Eq.\ (\ref{eq:jevaldist}).
It factorizes into the eigenvalue distribution
of $M-1$ dimensional Gaussian matrices $H'$ [again distributed according
to
Eq.\ (\ref{eq:hdist})],
and the term $\prod_{j=1}^{M-1}
|\omega_i|^\beta=|\det {H'}|^\beta$.

In the first step of our calculation, 
we use the Fourier representation of the $\delta$-functions
in Eq.\ (\ref{eq:pabdef})
and write
\begin{eqnarray}
&&P(A,B)\nonumber\\ &&\propto
\left\langle
\int\limits_{-\infty}^\infty
dx \,
\int\limits_{-\infty}^\infty dy\,
e^{ixA+iyB)}
\prod_{m=1}^{M-1}
\int\limits_{0}^\infty d|\alpha_m|^2 \, P(|\alpha_m|^2)
\right.
\nonumber
\\
&&
\left.
{}\times
\exp\left[-ix
\sum_{m=1}^{M-1}\frac{|\alpha_m|^2}{\omega_m}
-iy
\sum_{m=1}^{M-1}\frac{|\alpha_m|^2}{\omega_m^2}
\right]
\right\rangle
,
\end{eqnarray}
where the average refers to the variables $\{\omega_m\}$.
The integrals over $|\alpha_m|^2$ can be performed, resulting in
\begin{eqnarray}
&&P(A,B)\propto
\int dx \int  dy\, e^{ixA+iyB}
\nonumber
\\
&&{}\times\left\langle
\frac{\det H'^{2\beta}}{\det[H'^2+2iw(xH'+y)/\pi^2\beta]^{\beta/2}}
\right\rangle
,
\end{eqnarray}
where the average is now over the Gaussian ensemble of $H'$-matrices.
It is our goal to relate this average to autocorrelators
of the secular polynomial 
of Gaussian distributed random matrices, given
in Refs.\ \cite{andreev1995,kettemann}.

The determinant in the denominator can be expressed as a Gaussian
integral,
\begin{eqnarray}
&&P(A,B)\nonumber\\ && \propto
\int dx \int dy
\, e^{ixA+iyB}
 \int d {\bf z} \int dH' \,
\det H'^{2\beta}
\nonumber
\\
&&{}\times
\exp\left[-\frac{\beta \pi^2}{4M}\spur H'^2
-{\bf z}^\dagger \left(H'^2+\frac{2iw}{\beta\pi^2}(xH'+y)\right){\bf z}\right]
,
\nonumber
\\
\end{eqnarray}
where the $M-1$ dimensional vector $\bf z$ is real (complex) for $\beta=1$ (2).
Since our original expression did only depend on the eigenvalues of $H'$,
the formulation above is invariant under orthogonal (unitary) transformations
of $H'$, and we can choose a basis in which 
$\bf z$ points into the direction of the last basis vector (index $M-1$).
Let us denote the Hamiltonian in the block form
\begin{equation}
H'=\left(
\begin{array}{cc}
V & {\bf h} \\
{\bf h}^\dagger & g
\end{array}
\right)
.
\end{equation}
Here $V$ is a $(M-2)\times(M-2)$ matrix, $g$ a number, and $\bf
h$ a $(M-2)$ dimensional vector.
In this notation,
\begin{eqnarray}
P(A,B)&\propto&
\int dx \int dy\, e^{ixA+iyB}
\int d {\bf z}
\int dg\int dV\int d{\bf h}\,
\nonumber
\\
&&
{}\times
\det \left[V^{2\beta} (g-{\bf h}^\dagger V^{-1}{\bf
h})^{2\beta}\right]
\nonumber
\\
&&{}\times
\exp\left[-\frac{\beta \pi^2}{4M}(g^2+2|{\bf h}|^2+
\spur V^2)\right]
\nonumber
\\
&&
{}\times
\exp\left[
-|{\bf z}|^2 \left(g^2+|{\bf h}|^2 +\frac{2iw}{\beta\pi^2}(xg+y)\right)
\right]
.
\nonumber
\\
\end{eqnarray}
The integrals over $x$ and $y$ give
$\delta$-functions,
\begin{eqnarray}
&&P(A,B)\nonumber\\&&\propto
\int  d {\bf z}\int
dg\int dV\int d{\bf h}\,\det\left[ V^{2\beta} (g-{\bf h}^\dagger V^{-1}{\bf
h})^{2\beta}\right]
\nonumber
\\
&&{}\times
\exp\left[-\frac{\beta \pi^2}{4M}(g^2+2|{\bf h}|^2+
\spur V^2)
-|{\bf z}|^2 (g^2+|{\bf h}|^2)\right]
\nonumber
\\
&&
{}\times
\delta\left(A-gB\right)
\delta\left(B-2w|{\bf z}|^2/\beta\pi^2\right).
\end{eqnarray}
We then integrate over $g$ and ${\bf z}$,
\begin{eqnarray}
&&P(A,B)\nonumber\\ &&{} \propto
\int dV\, d{\bf h}\,\det\left[ V^{2\beta} \left(
\frac{A}{B}-{\bf h}^\dagger V^{-1}{\bf h}
\right)^{2\beta}\right]
B^{\frac{\beta}{2}(M-1)-2}
\nonumber
\\
&&{}\times
\exp\left[-\frac{\beta \pi^2}{4M}
\left(
2|{\bf h}|^2+
\spur V^2\right)
-\frac{\beta \pi^2 B}{2w} \left(\frac{A^2}{B^2}+|{\bf h}|^2
\right)\right]
.
\nonumber
\\
\label{eq:almostdone}
\end{eqnarray}
We already anticipated $B\gg 1/M$
and omitted in the exponent 
a term $-\beta \pi^2A^2/4MB^2$.

The integral over ${\bf h}$ can
be interpreted as an average over
 Gaussian random variables with variance
\begin{equation}
h^2\equiv \langle |h_i|^2\rangle=
\frac{1}{\pi^2}\frac{1}{B/w+1/M}\approx
\frac{w}{\pi^2 B}\left(1-\frac{w}{MB}\right).
\label{eq:varh}
\end{equation}
For the stochastic interpretation one
also has to supply the normalization constants  proportional to
\begin{equation}
h^{\beta(M-2)}=
\left(\frac{w}{\pi^2 B}\right)^{\beta(M-2)/2}
\exp\left[-\frac{\beta w}{2B}\right]. 
\end{equation}
The integral over $V$ is another Gaussian average, and thus
\begin{mathletters}
\begin{eqnarray}
&& P(A,B)\propto Q_\beta
B^{\frac{\beta}{2}-2}
\exp\left[-\frac{\beta w}{2B}\left(1+\frac{\pi^2A^2}{w^2}\right)\right]
,
\\
&&
Q_{\beta}=
\left\langle\det \left[V^{2\beta}\left(\frac{A}{B}
-{\bf h}^\dagger V^{-1}{\bf h} \right)^{2\beta} \right]\right\rangle
.
\end{eqnarray}
\end{mathletters}

After averaging over $\bf h$, one has now to consider for $\beta=1$
\begin{equation}
Q_1=\left\langle \det \left[
V^2\frac{A^2}{B^2}+ h^4 V^2\left[(\spur V^{-1})^2+2\spur V^{-2}
\right]
\right]
\right\rangle
,
\end{equation}
where only the even terms in $V$ have been kept.
The ratio of coefficients in this polynomial in $A/B$ 
can be calculated from the autocorrelator \cite{kettemann}
\begin{eqnarray}
G_1(\omega,\omega')&=&\frac{\langle \det (V+\omega)(V+\omega')\rangle}{
\langle \det V^2\rangle }
\nonumber\\
&=& \left. -\frac{3}{\pi^2 x}
\frac{d}{d x}
\frac{\sin \pi x}{\pi x}\right|_{x=\omega-\omega'}
\label{eq:g1}
\end{eqnarray}
of the secular
polynomial of Gaussian distributed real matrices $V$. This is achieved
by
expressing the products of traces and determinants through secular
coefficients, and these then as derivatives of the secular determinant,
\begin{mathletters}
\begin{eqnarray}
&&\frac{
\left\langle \det V^{2}(\spur V^{-1})^2\right\rangle 
}{\langle \det V^2\rangle }
=
\left.\frac{\partial^2}{\partial \omega\partial \omega'}
G_1(\omega,\omega')\right|_{\omega=\omega'=0}
\nonumber\\
&&\quad{}
=\left.-\frac{\partial^2}{\partial
\omega^2}G_1(\omega,0)\right|_{\omega=0}=\frac{\pi^2}{5},
\\
&&\frac{2\left\langle \det V^{2}(\spur V^{-2})
\right\rangle }{\langle \det V^2\rangle }=
-\left.4\frac{\partial^2}{\partial \omega^2}G_1(\omega,0)\right|_{\omega=0}.
\end{eqnarray}
\end{mathletters}
[We used the 
translational invariance of $G(\omega,\omega')$.]
Eqs.\ (\ref{eq:varh}) and (\ref{eq:g1}) yield
\begin{equation}
Q_1\propto \frac{A^2}{B^2}+\frac{w^2}{\pi^2 B^2}
.
\end{equation}

For $\beta=2$, the average over ${\bf h}$ yields
the expression
\begin{mathletters}
\begin{eqnarray}
&&Q_2\propto \frac{A^4}{B^4}+q_1 h^4 \frac{A^2}{B^2} +q_2 h^8,
\\
&&q_1=6\left\langle \det V^4 [(\spur V^{-1})^2+\spur
V^{-2}]\right\rangle,
\\
&&q_2=
\left\langle \det V^4
[(\spur V)^{-4}+6\spur V^{-2}(\spur V^{-1})^2
\right.
\nonumber
\\
&&
\left.{}
+8\spur V^{-1}\spur V^{-3}
+6\spur V^{-4} +3(\spur V^{-2})^2]\right\rangle
.
\end{eqnarray}
\end{mathletters}
The coefficients can now be computed from the
four-point correlator of the Gaussian unitary ensemble
\cite{andreev1995},
\begin{mathletters}
\begin{eqnarray}
&&G_2(\omega_1,\omega_2,\omega_3,\omega_4)
\nonumber
\\
&&=\frac{\langle\det(V+\omega_1)(V+\omega_2)(V+\omega_3)(V+\omega_4)\rangle}{
\langle\det V^4\rangle}
\nonumber
\\
&&=\frac{3}{2\pi^4}\left[
\frac{\cos\pi(\omega_1+\omega_2-\omega_3-\omega_4)}{
(\omega_1-\omega_3)(\omega_1-\omega_4)(\omega_2-\omega_3)(\omega_2-\omega_4)}
\right.
\nonumber
\\&&
{}
+
\frac{\cos\pi(\omega_1+\omega_3-\omega_2-\omega_4)}{
(\omega_1-\omega_2)(\omega_1-\omega_4)(\omega_3-\omega_2)(\omega_3-\omega_4)}
\nonumber
\\&&
\left.
{}
+
\frac{\cos\pi(\omega_1+\omega_4-\omega_3-\omega_2)}{
(\omega_1-\omega_3)(\omega_1-\omega_2)(\omega_4-\omega_3)(\omega_4-\omega_2)}
\right]
,
\\
&&G_2(\omega,0,0,0)=\frac{3}{\pi^3\omega^3}(\sin \pi \omega-\pi \omega \cos\pi \omega),
\\
&&G_2(\omega,\omega,0,0)=\frac{3}{2\pi^4\omega^4}(\cos 2\pi \omega-1+2\pi^2\omega^2).
\end{eqnarray}
\end{mathletters}
In this case
\begin{mathletters}
\begin{eqnarray}
q_1&=&
\left.\frac{\partial^2}{\partial \omega^2}\left[6 G_2(\omega,\omega,0,0)-
18 G_2(\omega,0,0,0)\right]
\right|_{\omega=0}
\nonumber
\\
&=&2\pi^2
,
\\
q_2&=&
\left. \frac{\partial^4}{\partial \omega^4}\left[10G_2(\omega,\omega,0,0)-
15G_2(\omega,0,0,0)\right]
\right|_{\omega=0}
\nonumber
\\
&=&\pi^2,
\end{eqnarray}
\end{mathletters}
which gives
\begin{eqnarray}
Q_2&\propto&Q_1^2
.
\end{eqnarray}

Collecting results we obtain Eq.\ (\ref{eq:pab}),
where we also included the normalization constant.

\section{Derivation of Eq.\ \lowercase{(\ref{eqs:meankfinal})} for
the mean Petermann factor}
\label{app:b}

The computation of the mean Petermann factor from the expression
(\ref{eq:genf})
is facilitated by the fact that it can be obtained from the same
generating function \cite{fyodorov:97a,sommers:88},
\begin{eqnarray}
&&\Psi(\omega_1,\omega_2,u_1,u_2,\veps)
\nonumber
\\
&&=\left\langle
\frac{\det[(\omega-\hmat)(\omega^*-\hmat^\dagger)-(u_1-i\veps)(u_2-i\veps)]}{
\det[(\omega-\hmat)(\omega^*-\hmat^\dagger)-(u_1+i\veps)(u_2+i\veps)]}
\right\rangle
,
\label{eq:zgen}
\end{eqnarray}
as the distribution function
\begin{eqnarray}
\rho(\omega)&=&\lim_{\veps\to 0^+}\left\langle
\spur
\frac{\varepsilon}{(\omega^*-\hmat^\dagger)(\omega-\hmat)+\varepsilon^2}
\right.
\nonumber
\\
&&\quad{}\times
\left.
\frac{\varepsilon}{(\omega-\hmat)(\omega^*-\hmat^\dagger)+\varepsilon^2}
\right\rangle
\end{eqnarray}
of poles in the complex
plane. (The distribution of poles is related to the distribution of
decay rates by
$P(\Gamma)=\frac{1}{2}\Delta\rho(\omega)|_{\omega=\Omega-i\Gamma/2}$.)
The relations are
\begin{eqnarray}
&&\pi \rho(\omega)
=\lim_{\veps\to 0^+}
\left(
\frac{\partial^2}{\partial \omega_2\partial\omega_2^*}
+\case{1}{2}\frac{\partial^2}{\partial \omega_2\partial\omega_1^*}
+\case{1}{2}\frac{\partial^2}{\partial \omega_1\partial\omega_2^*}
\right)
\nonumber \\
&&\left.\quad{}\times
\Psi(\omega_1,\omega_2,0,0,\veps)\right|_{\omega_1=\omega_2=\omega}
,
\\
&&M \pi \langle K \rangle_{\Omega,\Gamma}\rho(\omega)\nonumber\\
&&=-\lim_{\veps\to 0^+}\left.
\frac 14 \frac{\partial}{\partial u_1}
\frac{\partial}{\partial u_2} \Psi(\omega,\omega,u_1,u_2,\veps)
\right|_{u_1=u_2=0}
\label{eq:genf2}
.
\end{eqnarray}
Most of the analysis runs therefore in parallel with the calculation
of $\rho(\omega)$ in Ref.\ \cite{fyodorov:97a}.
We restrict ourselves to the case of broken time-reversal symmetry,
where the algebra is less involved. 

The ratio of determinants 
in Eq.\ (\ref{eq:zgen})
can be written as a superdeterminant, which in
turn can be expressed as a Gaussian integral over bosonic and fermionic
variables,
\begin{eqnarray}
\Psi(\omega,\omega,\veps,u_1,u_2)&=&
(-1)^M\langle \sdet^{-1}(A)\rangle
\nonumber\\
&=&
(-1)^M
\left\langle
\int d\pvec^\dagger \int d\pvec e^{i\pvec^\dagger A \pvec}
\right\rangle
.
\label{eq:susyint}
\end{eqnarray}
The matrix $A$ is
\begin{eqnarray}
A&=&
\left(
\begin{array}{cccc}
\omega-\hmat & 0 & i\veps+u_1 & 0 \\
0 & \omega-\hmat & 0 & -i\veps+u_1 \\
-i\veps-u_2 & 0 & -\omega^*+\hmat^\dagger & 0 \\
0 & -i\veps+u_2 & 0 & \omega^*-\hmat^\dagger
\end{array}
\right)
\nonumber
\\
&=&
(\Omega-H)\otimes \lmat+
i\left(\pi W^\dagger W-\frac{\Gamma}{2}\right)\otimes \sigmaz \lmat
\nonumber
\\
&&{}-i\veps\otimes\sigmax \lmat + \umat\sigmax\lmat
.
\end{eqnarray}
The vector $\pvec=\pvec_1\oplus\pvec_2\oplus\pvec_3\oplus\pvec_4$
is a $4M$-dimensional supervector consisting of two $M$-dimensional
bosonic entries $\pvec_{\alpha}$ with $\alpha=1$ and $3$,
supplemented by two $M$-dimensional
fermionic entries with $\alpha=2$ and $4$.
We encounter the  4-dimensional supermatrices $\lmat=\diag(1,1,-1,1)$,
$\umat=\diag(-u_1,u_1,-u_2,u_2)$, and ${\hat \sigma}_i=
\sigma_i\otimes \eins_2$, where
$\sigma_i$ are the usual Pauli matrices
[e.g.\ ${\hat\sigma}_z=\diag(1,1,-1,-1)$].

The linear appearance of $H$ in the exponent of Eq.\ (\ref{eq:susyint})
facilitates the ensemble average with the distribution function
(\ref{eq:hdist}),
since the integral over the independent components of $H$
factorizes, and each single integral is Gaussian.
The result is
\begin{mathletters}
\begin{eqnarray}
\left\langle
\exp[-i\pvec^\dagger H\otimes \lmat\pvec]
\right\rangle&=&\exp\left[-\frac{\mu^2M}{2}\str(\lmat \rmat)^2\right],
\nonumber
\\
\\
\rmat_{\alpha\beta}&=&\frac{1}{M}\pvec_{\alpha}\cdot\pvec_{\beta}^\dagger
.
\label{eq:rdef}
\end{eqnarray}
\end{mathletters}
The order of $\rmat$ in the exponent is reduced from quadratic to
linear by a
Hubbard-Stratonovich transformation,
based on the identity
\begin{eqnarray}
&&\exp\left[-\frac{\mu^2M}{2}\str(\lmat \rmat)^2\right]
\nonumber\\
\quad
&&=\int d\samat\,\exp\left[-M\str\left(\frac {\samat^2}{2}
-i\mu\samat\lmat \rmat\right)\right]
.
\end{eqnarray}
The integral over $\pvec$ and $\pvec^\dagger$ is again Gaussian
and results in 
\begin{mathletters}
\begin{eqnarray}
\Psi&=&\int d\samat\,
\exp\left[ -M\str\left(\frac {\samat^2}{2}+\ln \samat\right)\right]
\sdet^{-1}\left(1+C\right)
,
\nonumber\\
\label{eq:beforesp}
\\
C&=&\left(\Omega+
i\left(\pi W^\dagger W-\frac{\Gamma}{2}\right)
\otimes \sigmaz-i\veps\sigmax+\umat\sigmax
\right)\frac{1}{\mu \samat}
.
\nonumber
\\
\end{eqnarray}
\end{mathletters}

One now can write $\sdet^{-1} (1+C)=\exp[-\str\ln (1+C)]$
and expand the logarithm to first order
in $\Gamma$, $\veps$, and the source term $J$; in addition we set
$\Omega=0$ and pass
from the generating function to the
mean Petermann factor according to Eq.\ (\ref{eq:genf2}).
This gives
\begin{eqnarray}
&&M \pi \langle K\rangle_{\Omega,\Gamma}\rho(\omega)\nonumber\\
&&=-\frac{1}{4}\frac{\pi^2}{\Delta^2}
\int d\samat\,
\exp\Bigg[ -M\str\left(\frac {\samat^2}{2}+\ln \samat\right)
\nonumber
\\
&&{}
+i\frac{y}{2}\str\sigmaz\samat^{-1}
+i\frac{\veps'}{2}\str\sigmax\samat^{-1}
\Bigg]
\nonumber\\
&&{}\times
\mbox{tr}_{12}\sigma_x\samat^{-1}
\mbox{tr}_{34}\sigma_x\samat^{-1}
\prod_{n=1}^N\sdet^{-1}\left(\eins_4
+iw_n
\sigmaz\samat^{-1}\right).
\nonumber\\
\end{eqnarray}
The traces 
$\mbox{tr}_{ij}A=A_{ii}+A_{jj}$ operate only on the indicated subspaces.
We introduced
the rescaled variables $y=-2\pi\imagteil \omega/\Delta=\pi\Gamma/\Delta$ 
and $\veps'=2\pi\veps/\Delta$. In what follows we will write
$\veps$ instead of $\veps'$.

The condition $M\gg 1$ justifies a saddle-point approximation.
The main contribution to the preceding integral comes from points
for which the first part of the exponent is minimal,
that is from the solutions of
\begin{equation}
\frac{1}{\samat}+\samat=0 \quad \Leftrightarrow \quad \samat^2=-1.
\end{equation}
With $\samat=i\qmat$, the solutions fulfill $\qmat^2=1$. As inherited from the
definition of $\rmat$ in Eq.\ (\ref{eq:rdef}), $\qmat\lmat$ is a Hermitian matrix
and $\qmat=\tmat^{-1}\qmat_{\rm diag}\tmat$
can be diagonalized by a pseudounitary supermatrix
$\tmat\in U(1,1/2)$ (these matrices
fulfill $\tmat^\dagger\lmat \tmat=\lmat$).
The largest manifold
which respects the definiteness-requirements on $\qmat$
is obtained by the choice
$\qmat_{\rm diag}=\sigmaz$. However, rotations in the block $\alpha=1,3$
and in the block $\alpha=2,4$ leave  $\qmat$ invariant; the saddle-point
manifold is hence covered exactly once if we take the $\tmat$ matrices
from the coset space
$U(1,1/2)/U(1/1)\times U(1/1)$.

A convenient parameterization of the coset space has been given by Efetov
\cite{efetov}, 
\begin{mathletters}
\begin{eqnarray}
\tmat&=&
\left(
\begin{array}{cc}
U^{-1}&0\\
0&V^{-1}
\end{array}
\right)
\exp
\left(
\begin{array}{cc}
0&\frac{1}{2}\diag(\theta_1,i\theta_2)\\
\frac{1}{2}\diag(\theta_1,i\theta_2)&0
\end{array}
\right)
\nonumber
\\
&&{}\times
\left(
\begin{array}{cc}
U&0\\
0&V
\end{array}
\right)
,
\end{eqnarray}
\begin{equation}
U=
\left(
\begin{array}{cc}
e^{i\phi_1}&0\\
0&e^{i\phi_2}
\end{array}
\right)
\left(
\begin{array}{cc}
1+\rho\rho^*/2&\rho\\
\rho^*&1+\rho^*\rho/2
\end{array}
\right)
,
\end{equation}
\begin{equation}
V=\left(
\begin{array}{cc}
1-\sigma\sigma^*/2&i\sigma\\
i\sigma^*&1-\sigma^*\sigma/2
\end{array}
\right)
,
\end{equation}
\end{mathletters}
with bosonic variables $\theta_1$, $\theta_2$, $\phi_1$, and $\phi_2$,
and fermionic
variables $\rho$, $\rho^*$, $\sigma$, and $\sigma^*$.
We introduce $\lambda_1=\cosh\theta_1$ and $\lambda_2=\cos\theta_2$.
In this parameterization 
\begin{mathletters}
\begin{eqnarray}
&&\str\sigmaz\qmat
=2(\lambda_1-\lambda_2),
\\
&&\str\sigmax\qmat=
-\sinh \theta_1 e^{i\phi_1}
[(1+\rho\rho^*/2)(1-\sigma\sigma^*/2)-i\rho\sigma^*]
\nonumber
\\
&&{}
+\sinh \theta_1 e^{-i\phi_1}
[(1+\rho\rho^*/2)(1-\sigma\sigma^*/2)-i\sigma\rho^*]
\nonumber
\\
&&{}
+i\sin \theta_2 e^{i\phi_2}
[(1+\rho^*\rho/2)(1-\sigma^*\sigma/2)-i\rho^*\sigma]
\nonumber
\\
&&{}
-i\sin \theta_2 e^{-i\phi_2}
[(1+\rho^*\rho/2)(1-\sigma^*\sigma/2)-i\sigma^*\rho]
,
\\
&&\mbox{tr}_{12}\sigma_x\qmat=
-\sinh \theta_1 e^{i\phi_1}
[(1+\rho\rho^*/2)(1-\sigma\sigma^*/2)-i\sigma^*\rho]
\nonumber
\\
&&{}
-i\sin \theta_2 e^{i\phi_2}
[(1+\rho^*\rho/2)(1-\sigma^*\sigma/2)-i\sigma\rho^*]
,
\\
&&\mbox{tr}_{34}\sigma_x\qmat=
\sinh \theta_1 e^{-i\phi_1}
[(1+\rho\rho^*/2)(1-\sigma\sigma^*/2)-i\rho^*\sigma]
\nonumber
\\
&&{}
+i\sin \theta_2 e^{-i\phi_2}
[(1+\rho^*\rho/2)(1-\sigma^*\sigma/2)-i\rho\sigma^*]
,
\\
&&
\sdet^{-1}[ \eins_4+w_n\sigmaz\qmat]=
\frac{g_n+\lambda_2}
{g_n+\lambda_1}
.
\end{eqnarray}
\end{mathletters}
The integration measure is
\begin{eqnarray}
d\qmat
&=&
\frac{d\lambda_1\,d\lambda_2\,
d\phi_1\,d\phi_2\,d\rho^*\,d\rho\,di\sigma^*\,di\sigma}{
(2\pi)^2(\lambda_1-\lambda_2)^2}
.
\end{eqnarray}

In order to integrate over the fermionic variables we have to expand
in these quantities and only keep the term in which all four 
variables appear linearly. The angle $\phi_2$ appears in the
pre-exponential factor as well as in the exponential term
$\exp(-\veps \sin\theta_2\sin\phi_2)$.
We expand the exponential and integrate
over $\phi_2$. Only terms
of order $\veps^n\sinh^m\theta_1$ with $n\le m$ survive the limit
$\veps\to 0$. We discard all other terms and obtain
\begin{mathletters}
\begin{eqnarray}
&&-4\frac{\Delta^2}{\pi^2}\langle K\rangle_{\Omega,\Gamma}\rho(\omega)
\nonumber
\\
&&{}=\lim_{\veps\to 0}
\int_1^\infty d\lambda_1\int_{-1}^1
d\lambda_2\,\frac{1}{(\lambda_1-\lambda_2)^2}\int_0^{2\pi}
\frac{d\phi_1}{2\pi} D
\nonumber
\\
&&{}\times
\exp\left[-i\veps\sqrt{\lambda_1^2-1}\sin\phi_1+y(\lambda_1-\lambda_2)
\right]
\prod_{n=1}^N\frac{g_n+\lambda_2}{g_n+\lambda_1}
,
\nonumber
\\
\label{eq:zwlang}
\\
&&D=-i\veps\sinh\theta_1\sin\phi_1(2\sin^2\theta_2+\case{9}{4}\sinh^2\theta_1)
\nonumber
\\
&&{}
+\frac{\veps^2}{4}\sinh^2\theta_1[
\sinh^2\theta_1\cos^2\phi_1
\nonumber
\\
&&\quad{}
-(3\cos^2\phi_1+5\sin^2\phi_1) \sin^2\theta_2
]
\nonumber
\\
&&{}
+i\veps^3\sinh^3\theta_1\sin\phi_1\sin^2\theta_2
(-\case{9}{16}\sin^2\phi_1-\case{13}{16}\cos^2\phi_1)
\nonumber
\\
&&{}
+\case{1}{16}\veps^4\sinh^4\theta_1\sin^2\phi_1\sin^2\theta_2
.
\end{eqnarray}
\end{mathletters}
It is convenient to bring the factor $D$ into a form which involves
$\phi_1$ only in the combination $z_1=-i\sinh\theta_1\sin\phi_1$,
because such terms can be expressed as derivatives
with respect to $\veps$ of the exponential
$\exp(\veps z_1)$ appearing in Eq.\ (\ref{eq:zwlang}).
This goal can be achieved by integrating by parts all terms that 
involve $\cos\phi_1$. Effectively this amounts to the substitutions
$\veps\sinh\theta_1\sin\phi_1\cos^2\phi_1\to i(\sin^2\phi_1-\cos^2\phi_1)$
and $\veps\sinh\theta_1\cos^2\phi_1\to i\sin\phi_1$, 
resulting in 
\begin{eqnarray}
D&=&\veps z_1(\case 7 2 \sin^2\theta_2+2\sinh^2\theta_1)+
\case{1}{2}\veps^2 z_1^2\sin^2\theta_2
\nonumber
\\
&&{}-\case{1}{2}\veps^3 z_1^3\sin^2\theta_2
.
\label{eq:d}
\end{eqnarray}

Mathematically these expressions are 
quite similar to those obtained for the decay-rate
distribution in Ref.~\cite{fyodorov:97a}.
By a simple substitution rule that relates to each other the 
terms of different order in $\veps$, we now rewrite
$D$ in a way that allows to 
make direct contact to Ref.~\cite{fyodorov:97a}, yielding a
result in terms of the two functions ${\cal F}_{1,2}$ given in
Eq.\ (\ref{eq:pgue}):
As in Ref.~\cite{fyodorov:97a} we express the factors $(g_n+\lambda_1)^{-1}$
as an integral of exponential functions
\begin{equation}
\frac{1}{g_n+\lambda_1}=\int_0^\infty ds_n\,\exp[-s_n(g_n+\lambda_1)]
.
\end{equation}
We also write
$(\lambda_1-\lambda_2)^2=\int_0^\infty
dx\,x\exp[-x(\lambda_1-\lambda_2)]$.
Then the integrations over $\theta_1$ and $\phi_1$ can be performed, 
and $\veps$ only appears in a factor
\begin{equation}
\Phi(\veps,y')=\frac{\exp[-\sqrt{\veps^2+y^2}]}{\sqrt{\veps^2+y'^2}},
\end{equation}
with $y'=y-x-\sum_n s_n$.
The limiting value for $\veps\to 0$ of the derivatives 
\begin{equation}
\veps^n\frac{\partial^n}{\partial \veps^n}
\Phi(\veps,y')=C_n\delta(y'),\quad C_1=-C_2=C_3/2=-2,
\end{equation}
amounts in Eq.\ (\ref{eq:d})
to the substitutions $\veps^3 z_1^3\to 2\veps z_1$ and
$\veps^2 z_1^2\to -\veps z_1$, which gives
$D=2\veps z_1(\lambda_1^2-\lambda_2^2)$.
As a result, we obtain
\begin{eqnarray}
&&\left[\frac{\Delta}{\pi}
P(\Gamma)\right]\langle K \rangle_{\Omega,K}=
I_0(\pi\Gamma/\Delta)+2I_1(\pi\Gamma/\Delta)
,
\\
&&I_\nu(y)=-\frac{1}{4}\lim_{\veps\to 0}
\veps\frac{\partial}{\partial \veps}\int_1^\infty d\lambda_1\int_{-1}^1
d\lambda_2\frac{\lambda_2^\nu}{(\lambda_1-\lambda_2)^\nu}
\nonumber\\
&&{}\times
J_0(\veps\sqrt{\lambda_1^2-1})
\exp[y(\lambda_1-\lambda_2)]\prod_{n=1}^N
\frac{g_n+\lambda_2}{g_n+\lambda_1}
,
\end{eqnarray}
where $J_0$ is a Bessel function.
By comparing expressions with Ref.~\cite{fyodorov:97a}, we recognize
that $I_0(y)={\cal F}_1(y){\cal F}_2(y)=
(\Delta /\pi) P(\Gamma=\Delta y/\pi)$ [cf.\ Eq.\ (\ref{eq:pgue})], while
\begin{equation}
I_1(y)=-\int_0^{y}dy'\,
{\cal F}_1(y')\frac{\partial}{\partial y'}
{\cal F}_2(y')
.
\end{equation}
This concludes the derivation of the final result
(\ref{eqs:meankfinal}).

\end{document}